\pdfoutput=1
\documentclass{JINST}
\let\ifpdf\relax

\usepackage{subfig}

\title{An adjustable focusing system for a 2~MeV H\textsuperscript{--} ion beam line based on permanent magnet quadrupoles}

\author{M.~Nirkko$^a$\thanks{Corresponding author.} , S.~Braccini$^a$, A.~Ereditato$^a$, I.~Kreslo$^a$, P.~Scampoli$^{a,b}$ and M.~Weber$^a$\\\\
\llap{$^a$}Albert Einstein Center for Fundamental Physics, \\
Laboratory for High Energy Physics, University of Bern, \\
Sidlerstrasse 5, CH-3012 Bern, Switzerland \\
E-mail: \email{martti.nirkko@lhep.unibe.ch}\\
\llap{$^b$}Dipartimento di Scienze Fisiche, Universit\`{a} di Napoli Federico II, \\
Complesso Universitario di Monte S. Angelo, I-80126 Napoli, Italy}

\abstract{A compact adjustable focusing system for a 2 MeV H\textsuperscript{--} RFQ Linac is designed, constructed and tested based on four permanent magnet quadrupoles (PMQ). A PMQ model is realised using finite element simulations, providing an integrated field gradient of $2.35\;\mathrm{T}$ with a maximal field gradient of $57\;\mathrm{T \, m^{-1}}$. A prototype is constructed and the magnetic field is measured, demonstrating good agreement with the simulation. Particle track simulations provide initial values for the quadrupole positions. Accordingly, four PMQs are constructed and assembled on the beam line, their positions are then tuned to obtain a minimal beam spot size of $(1.2 \times 2.2)\;\mathrm{mm^2}$ on target. This paper describes an adjustable PMQ beam line for an external ion beam. The novel compact design based on commercially available NdFeB magnets allows high flexibility for ion beam applications.}


\begin{document}

\clearpage{}
\section{Introduction}
Magnetic quadrupoles are commonly used for focusing charged particle beams in accelerators. Mostly, electromagnets are used to achieve magnetic fields of sufficient strength. These require electricity and cooling, and are usually the best choice for high energy beam lines. However, in recent years developments in the field of permanent magnets have allowed for the use of permanent magnet multipoles in beam lines \cite{pmq-system}.

In this paper we report of the design and construction of a focusing system for a 2 MeV H\textsuperscript{--} ion beam using two FODO doublets, consisting of two PMQs each. This high energy beam transport (HEBT) configuration is compact, cost-efficient and easily adjustable in order to fit the needs of different experimental applications.
For fixed target experimental activities, a focused beam is desired in order to obtain higher interaction rates on a target. Such a setup enables applications in material analysis such as PIGE/PIXE \cite{pige1, pige2, peterson} and GRNA \cite{grna1, grna2, grna4}.
Another interesting option is the use of the envisioned PMQ system for beam transfer lines (BTL), where minimal beam divergence is required instead of minimal beam size. The scope of this work is to achieve small spot sizes for fixed target experiments, while maintaining maximum flexibility of the beam line.

\clearpage{}
\section{Materials and methods}

\subsection{The RFQ accelerator}
The 2 MeV RFQ Linac used here was originally conceived for the calibration of the BGO calorimeter of the L3 experiment at LEP \cite{rfq1,rfq2,rfq3}. The RFQ was decommissioned in 2000 and stored at CERN. In 2009, it was recommissioned and put into operation at the University of Bern. The accelerator provides a pulsed 1.92 MeV H\textsuperscript{--} ion beam at a repetition rate in the range of (10--150) Hz. The peak current and the pulse width are adjustable in the range (1--5) mA and (1--25) $\mathrm{\mu m}$, respectively. In the extraction beam line, beam cross sections between $(5\times5)\,\mathrm{mm^2}$ and $(50\times50)\,\mathrm{mm^2}$ can be obtained. 

\subsection{Beam emittance}
\begin{figure}[!ht]
	\centering
	\subfloat[ ]{
	\includegraphics[width=0.58\textwidth, trim=0.5cm 0cm 0.5cm 0cm, clip=true]{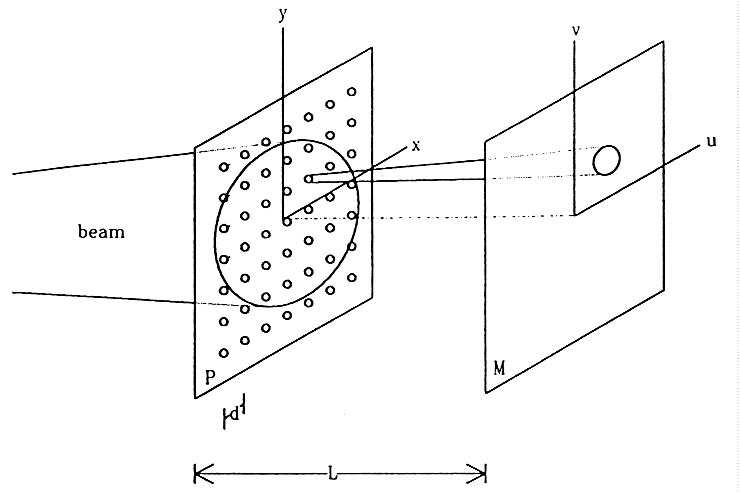}
	\label{ppot_scheme}}
	\subfloat[ ]{
	\includegraphics[width=0.38\textwidth, trim=3.5cm 2cm 5cm 1cm, clip=true]{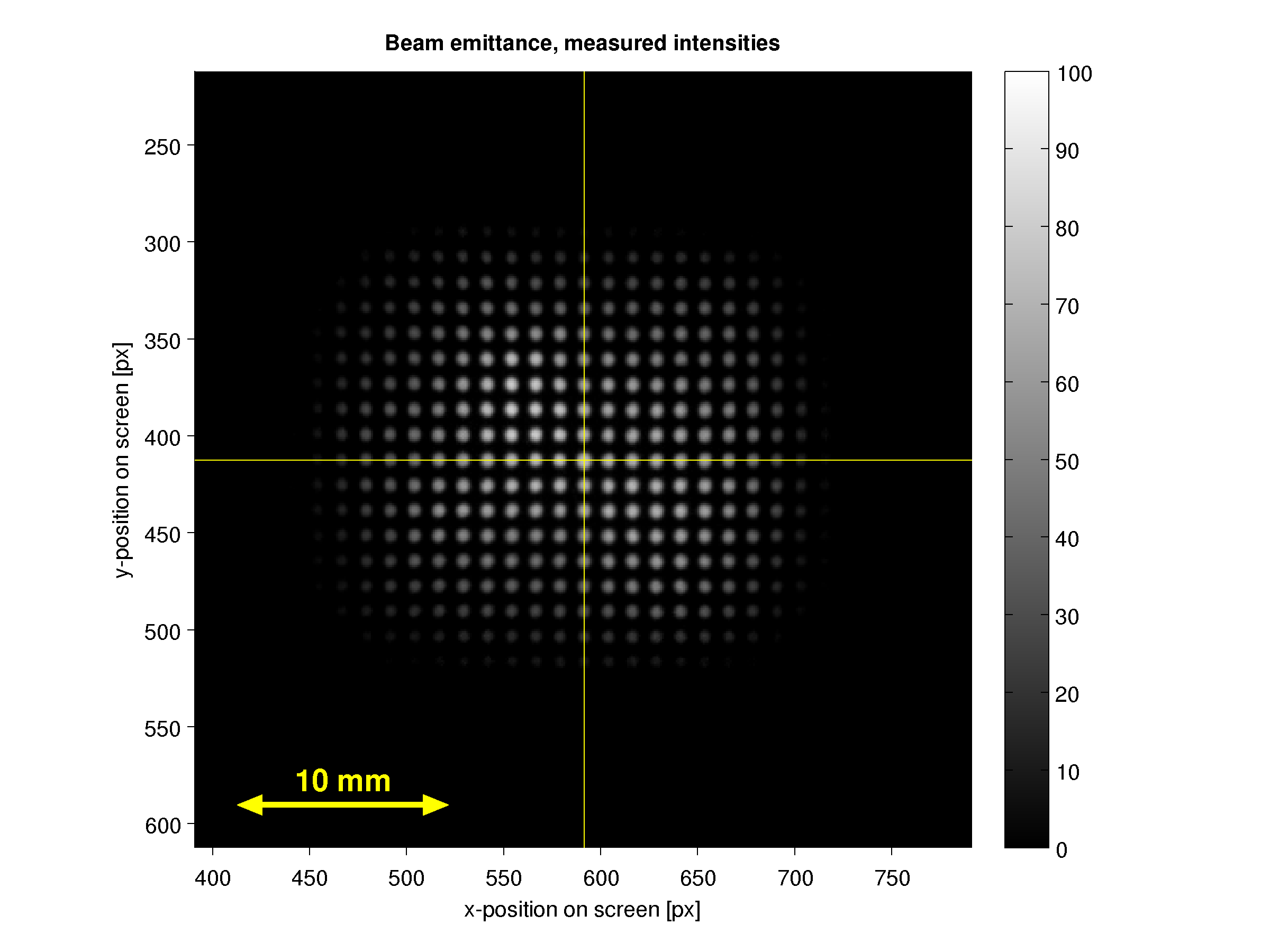}
	\label{ppot_data}}
	\caption{(a) Scheme of the ``pepper-pot'' method {\cite{emit1}} and (b) measured 2D dataset.}
	\label{pepper-pot}
\end{figure}

\begin{figure}[!ht]
	\centering
	\includegraphics[width=\textwidth, trim=0.5cm 0.5cm 0.5cm 1.3cm, clip=true]{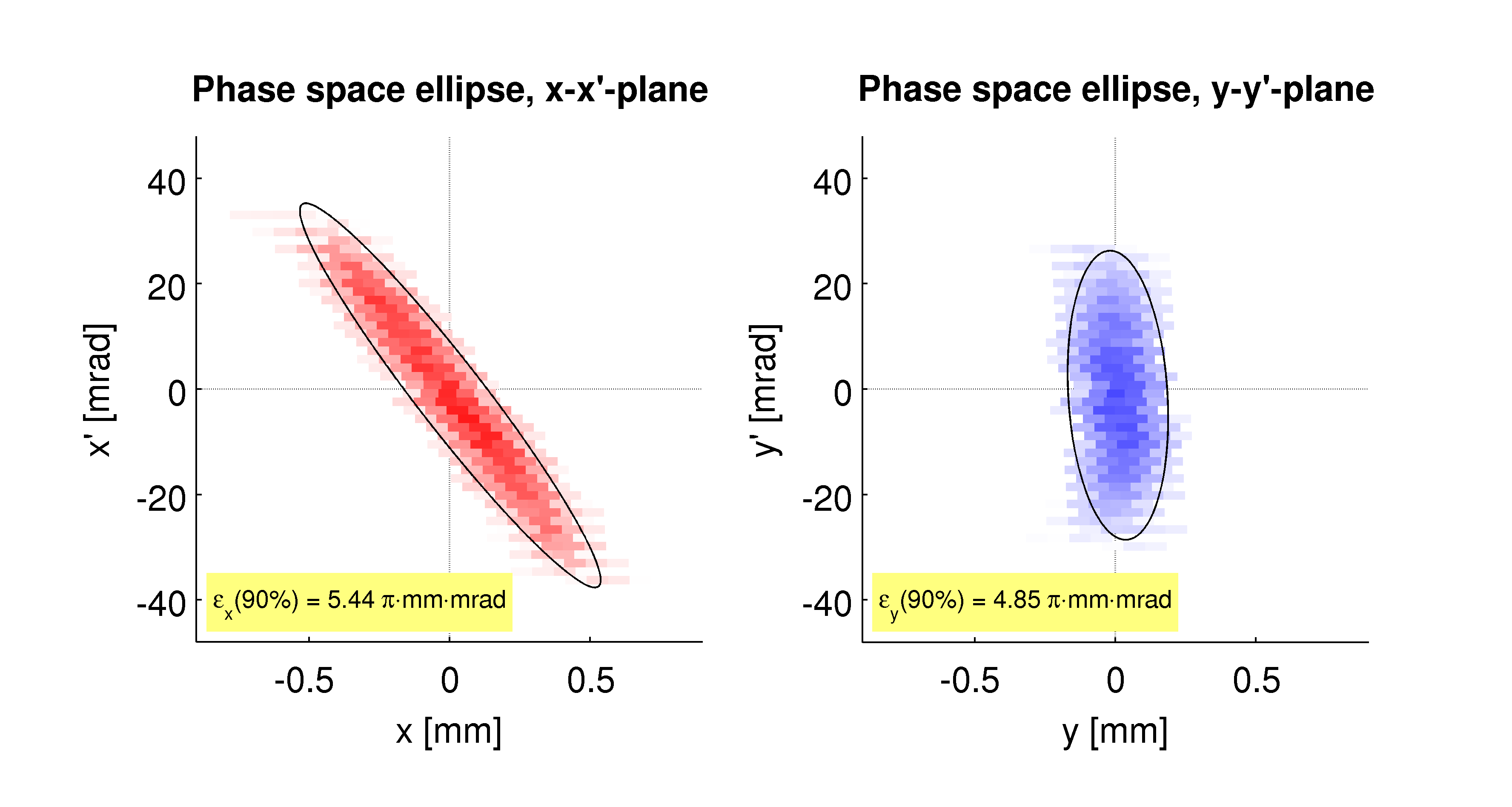}
	\caption{Measured transverse emittances in horizontal (left) and vertical (right) trace space, drifted back to the RFQ output plane.}
	\label{emittance}
\end{figure}

The nominal value for the normalised transverse output emittance (90\% of the full beam) of the RFQ Linac is $\mathrm{<0.6\,\pi\,mm\,mrad}$, given by the manufacturer \cite{accsys}. This corresponds to a geometric output emittance (90\%) of $\mathrm{<9.4\,\pi\,mm\,mrad}$. As the RFQ was completely dismantled and recommissioned several years later, it was necessary to revise this figure. An emittance measurement using the ``pepper-pot'' technique was conducted {\cite{emit1}}. A stainless steel hole-plate is placed in the beam line, thereby splitting the beam into several beamlets of equal size, which are equally separated by the distance $d$ between two holes. At a certain distance $L$ downstream of the plate, a beam monitor $M$ is placed, which measures the position and intensities of these beamlets. A scheme of this setup is shown in Fig.~\ref{ppot_scheme}.
In this paper, a stainless steel hole-plate of thickness 0.5 mm with a $25\times25$ matrix of holes was used, positioned at a distance $z_0=310$~mm from the RFQ output plane. The hole diameter was 0.5 mm, with a distance between holes of 1 mm. The monitor consisted of a 1 mm thick glass plate placed at $L=62$~mm downstream of the hole-plate, producing scintillation light as a result of the incoming beamlets. The amount of scintillation light produced corresponds to the intensity of the beamlets and was measured using a linear-response CCD camera. This 2D dataset was then processed in the following way \cite{emit1, emit2}: After subtracting the camera background from the image dataset, a small correction on the rotation of the camera was done using bilinear interpolation. Furthermore, noise was subtracted and false signals removed (e.g. reflection of light on the edges of the glass scintillator). For each beamlet, the intensities $I(x)$ and $I(y)$ through the centre of hole-plate (Fig.~\ref{ppot_data}) were then determined and their derivatives $dI(x)/dx$ and $dI(y)/dy$ calculated. The maxima and minima of the derivatives indicate the projected positions of the edges of each hole on the monitor. The points in trace space $(x,x',y,y')$ at the hole-plate (Fig.~\ref{ppot_scheme}) were then obtained as follows:
\begin{enumerate}
	\item The angles $u'$, $v'$ are calculated using the projected edges of the hole-plate on the screen.
	\item The positions $u$, $v$ are calculated using the obtained values within a few pixels around the found maxima and minima.
	\item Coordinates at the monitor $(u,v)$ are drifted upstream to coordinates at the hole-plate $(x,y)$.
\end{enumerate}
The centre hole on the hole plate was designed to be 20\% larger for reference, thus it needed to be normalised in the dataset. The points in trace space were then extrapolated upstream to the RFQ output plane: $x_0 = x_1 - z_0 \cdot x_1'\,$, $\;x_0' = x_1'$. Finally, ellipses were fitted around the data points in trace space, containing 90\% of the particles (Fig.~\ref{emittance}).
Results show that at current settings, the transverse output emittances (90\%) are $\mathrm{5.4\,\pi\,mm\,mrad}$ and $\mathrm{4.9\,\pi\,mm\,mrad}$ for the $x$ and $y$ coordinates respectively, a relatively small figure with respect to the value given by the manufacturer. These measured values were used as input for the beam line simulations reported in this paper.

\subsection{Design of the PMQs}
\begin{figure}[!ht]
	\centering
	\subfloat[ ]{
	\includegraphics[width=0.35\textwidth, trim=0cm 0cm 0cm 0cm, clip=true]{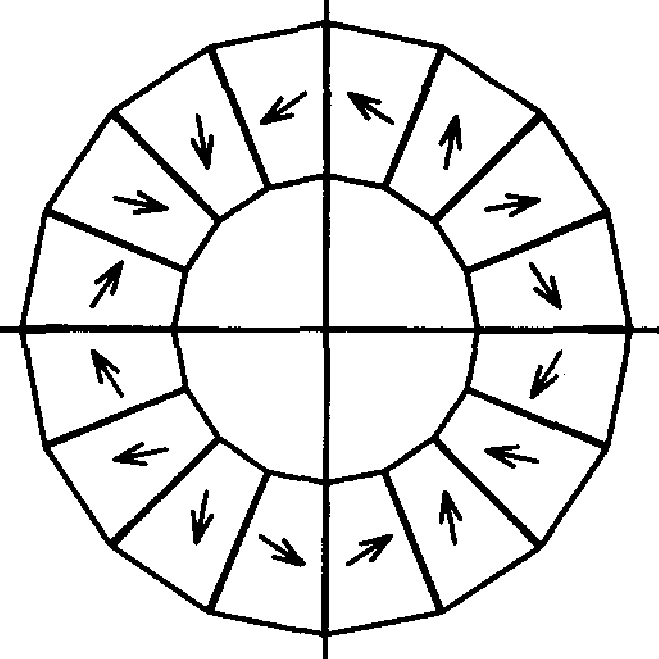}
	\label{halbach_16p}}
	\qquad
	\subfloat[ ]{
	\includegraphics[width=0.35\textwidth, trim=0.5cm 0.3cm 0.5cm 1cm, clip=true]{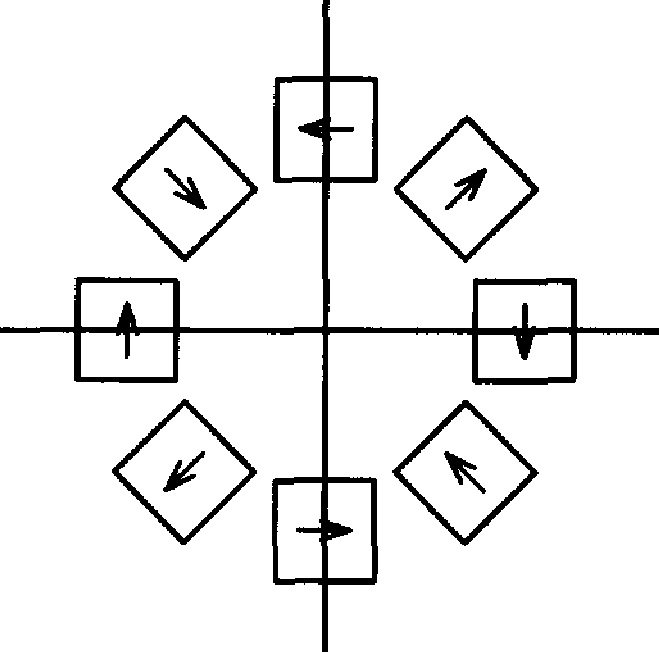}
	\label{halbach_8p}}
	\caption{Scheme of cross sections through Halbach cylinders using (a) 16 trapezoidal magnets and (b) 8 square block magnets, including direction of magnetisation \cite{halbach2}.}
	\label{halbach_cylinder}
\end{figure}

A Halbach array is an array of permanent magnets designed to augment the magnetic field on one side of the array, whereas cancelling it on the other. Such an array can be arranged into a ring, which is often called Halbach cylinder. This arrangement of permanent magnets creates a magnetic multipole, depending on their direction of magnetisation. Typical designs use 16 trapezoidal permanent magnets with a residual magnetisation $B_r$ to create a ``ring'' with an inner radius $r_i$ and an outer radius $r_o$. The resulting field inside such a quadrupole (distance from axis $r<r_i$) is 
\begin{equation} 
    B(r) = 2B_r\cdot r\left(\frac{1}{r_i}-\frac{1}{r_o}\right)K \; ,
    \label{halbach_formula}
\end{equation} 
where $K$ is a constant depending on the number of magnets $M$ used to create the PMQ. For $M=16$, $K=0.94$ and for $M=8$, $K=0.77$ \cite{halbach0}. The former design is shown in Fig.~\ref{halbach_16p}.

In this paper, a magnetic quadrupole using eight identical square permanent magnets as in Fig.~\ref{halbach_8p} is used. Criteria for choosing an optimal design included high field gradients, commercial availability of the magnets, and cost. The result is a compact, light, highly flexible PMQ which is easily disassembled and reassembled, and cost effective.

\begin{figure}[!ht]
	\centering
	\includegraphics[width=0.4\textwidth, trim=3cm 2.5cm 5cm 2.8cm, clip=true]{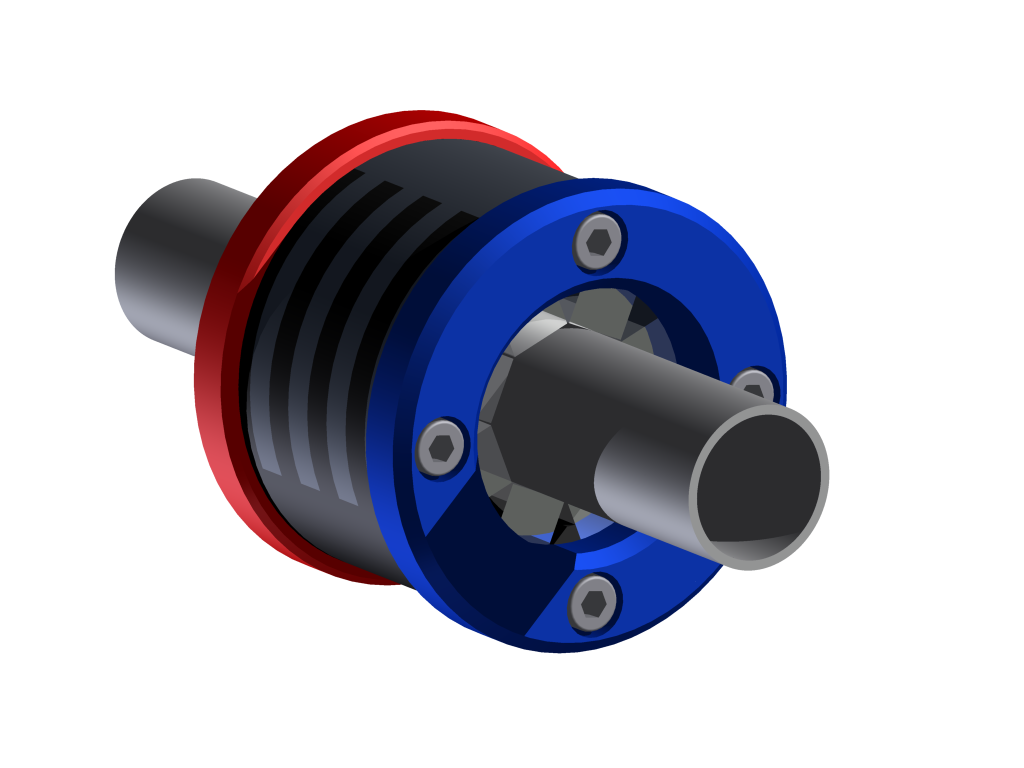}
	\caption{3-D rendering of designed PMQ. The square magnets are visible at the centre, wrapped around the beam pipe and held in place by the aluminium frame.}
	\label{pmq_tech}
\end{figure}

\begin{figure}[!ht]
	\centering
	\subfloat[ ]{
	\includegraphics[width=0.48\textwidth, trim=4.0cm 3.0cm 4.0cm 1.0cm, clip=true]{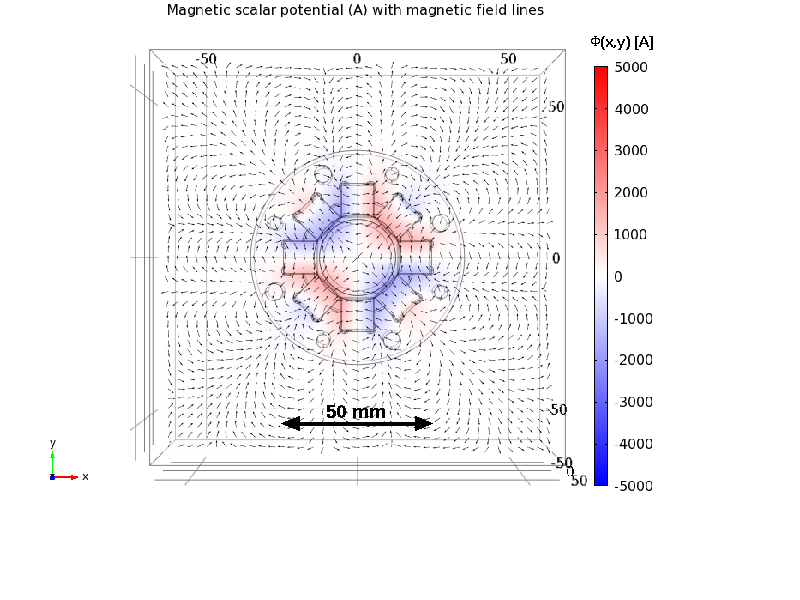}
	\label{pmq3_Axy}}
	\subfloat[ ]{
	\includegraphics[width=0.48\textwidth, trim=4.0cm 3.0cm 4.0cm 1.0cm, clip=true]{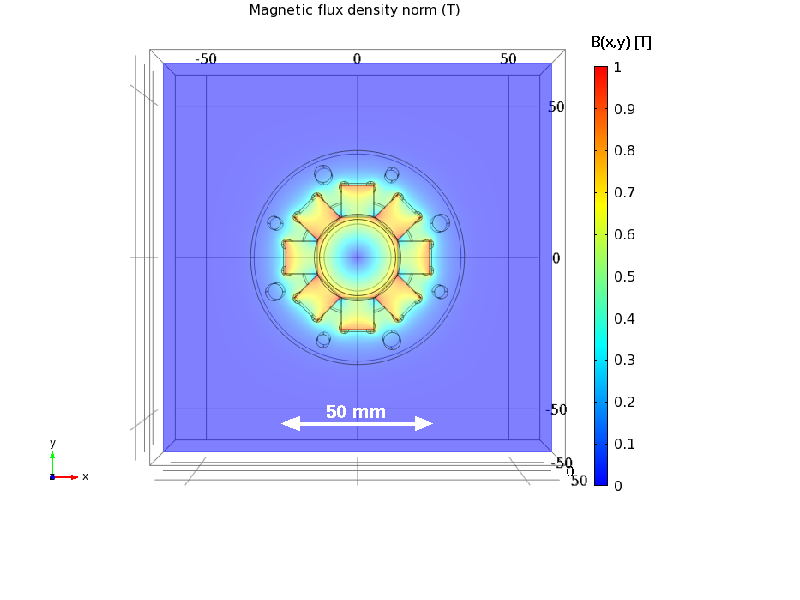}
	\label{pmq3_Bxy}}
	\caption{Simulated magnetic field at transversal cross section through longitudinal centre of designed PMQ. (a) Scalar magnetic potential with magnetic field vectors pointing from North (red) to South (blue). (b) Magnetic field strength with color scale from 0 (blue) to 1 T (red).}
	\label{pmq3_fields}
\end{figure}

Beam simulations of the envisioned setup were conducted to determine approximate requirements for the PMQs, resulting in field gradients of the order of 60 T/m and a PMQ length of 40 mm. Accordingly, the design implemented contains eight cuboid nickel-plated NdFeB magnets of dimensions $(40\times 10\times 10)\;\mathrm{mm^3}$ with a residual magnetisation of 1.3 T (grade N42) \cite{supermagnete}. These are supported by a metal frame on the outside and a stainless steel pipe ({\O}~=~25~mm) on the inside. The magnets are kept in place by a lid on each end, attached with four stainless steel screws. A 3-D rendering of the PMQ is shown in Fig.~\ref{pmq_tech}.
Materials with low relative magnetic permeabilities $\mu_r = \mu/\mu_0$ had to be chosen for the supporting frame, as materials with high $\mu_r$ would weaken the field at the centre of the PMQ. Due to its easy workability and a relative permeability of $\mu_r=1.000022$ \cite{mu-alu}, aluminium was chosen for the frame in the final implementation. According to Eq.~\ref{halbach_formula}, the magnetic field gradient $G = \frac{\partial B}{\partial r}$ at the longitudinal centre of an 8-piece PMQ should equal 71.2 T/m. As the magnets are not trapezoidal, this value is expected to be slightly smaller. The generated magnetic field of the described design was therefore simulated with Comsol\footnote{COMSOL Multiphysics\textsuperscript{\textregistered} is an engineering, design, and finite element analysis software environment for the modeling and simulation of any physics-based system. \href{http://www.comsol.com/}{www.comsol.com}} and is shown in Fig.~\ref{pmq3_fields}. The maximal field gradient was found to be 57.0 T/m, as expected.

\begin{figure}[!ht]
	\centering
	\includegraphics[width=0.75\textwidth, trim=0.5cm 0.2cm 0.5cm 0.2cm, clip=true]{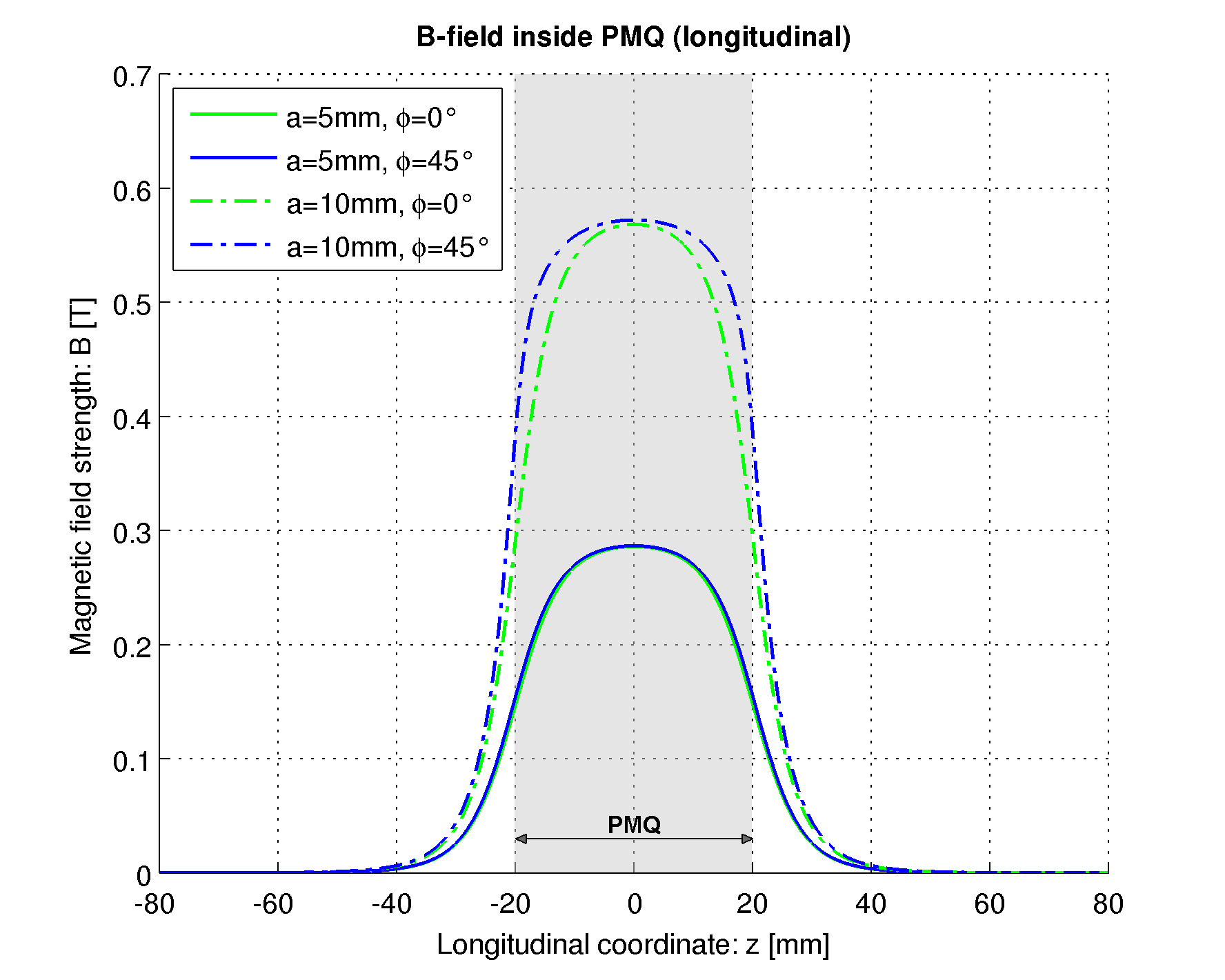}
	\caption{Simulated magnetic fields along longitudinal coordinate of designed PMQ.}
	\label{B_longitudinal}
\end{figure}

Concerning the longitudinal field component, fringe fields must be taken into account. Therefore, the longitudinal field at a radius $r=5\,\mathrm{mm}$ (Fig.~\ref{B_longitudinal}) was extracted from the 3D simulations and used as input for later particle track simulations. The integrated field gradient was found to be $\int G(z)\,dz = \mathrm{2.35\;T}$.

\subsection{Particle transport simulation}
\begin{figure}[!ht]
	\centering
	\includegraphics[width=\textwidth, trim=1.9cm 1.3cm 1.4cm 0.9cm, clip=true]{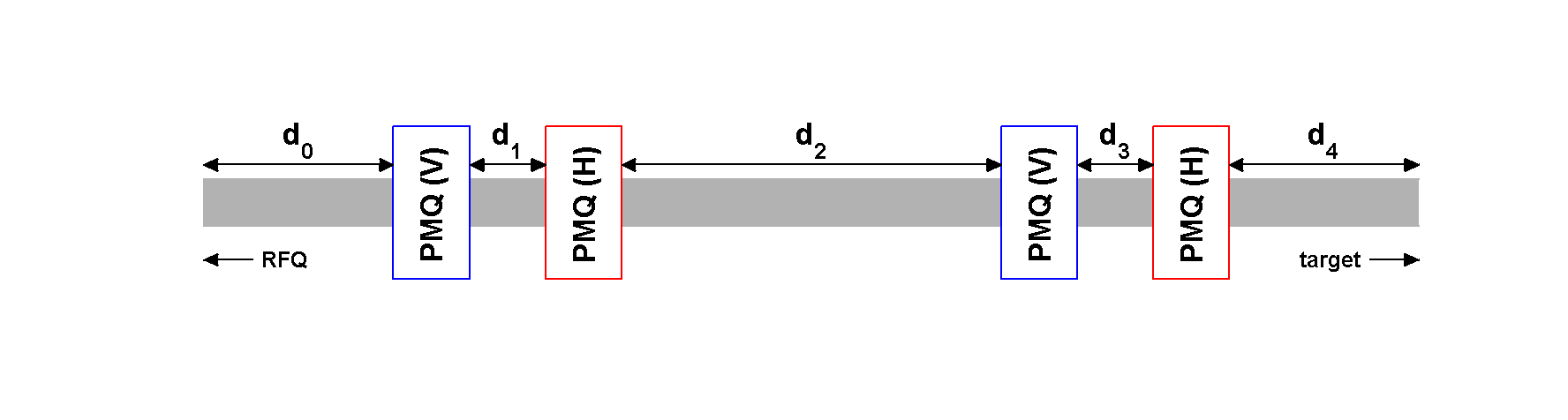}
	\caption{Scheme for intended HEBT beam line, consisting of two FODO doublets.}
	\label{hebt_example}
\end{figure}

For the beam line, a lattice of two FODO doublets was realised by four PMQs. Due to the focusing strength being constant for a given PMQ, the necessary degrees of freedom for beam steering must be obtained by adjusting the drift distances between elements $d_i$. This is shown schematically in Fig.~\ref{hebt_example}.
An initial parameter set of drift distances $\mathrm{[d_0\,d_1\,d_2\,d_3\,d_4]}$ for obtaining minimal spot sizes on target was obtained from a simulation of the beam line.

For particles entering the beam line at 1.92 MeV, calculations of single particle tracks were made using the simplified case of decoupled 2x2-matrices \cite{carey, hinterberger}. This simplification is valid assuming a short beam line with $L<1$~m in which the quadrupoles are ideally aligned, and dispersion effects and transverse coupling are therefore negligible. The simulation was written in Matlab\footnote{MATLAB\textsuperscript{\textregistered} is a high-level technical computing language and interactive environment for algorithm development, data visualization, data analysis, and numerical computation. \href{http://www.mathworks.com/}{www.mathworks.com}}, with which a Monte-Carlo simulation (MCS) using $10^3$ particles was run for many different parameter sets $[d_0\,d_1\,d_2\,d_3\,d_4]$. As input for the RFQ beam, particles with random trace space coordinates $(x,x',y,y')$, equally distributed within the measured emittance (Fig.~\ref{emittance}), were used. A magnetic field map for the designed PMQ was obtained from Comsol simulations and used to estimate the focusing factor $\kappa$ for each z-coordinate with step sizes $dz=1$~mm, taking into account fringe fields (Fig.~\ref{B_longitudinal}). The transfer matrix for the beam line was then obtained by sequentially multiplying each matrix for a thick quadrupole lens of thickness $dz$. It is to be noted that space charge effects were not taken into account.

\begin{figure}[!ht]
	\centering
	\includegraphics[width=\textwidth, trim=0cm 0cm 0cm 0cm, clip=true]{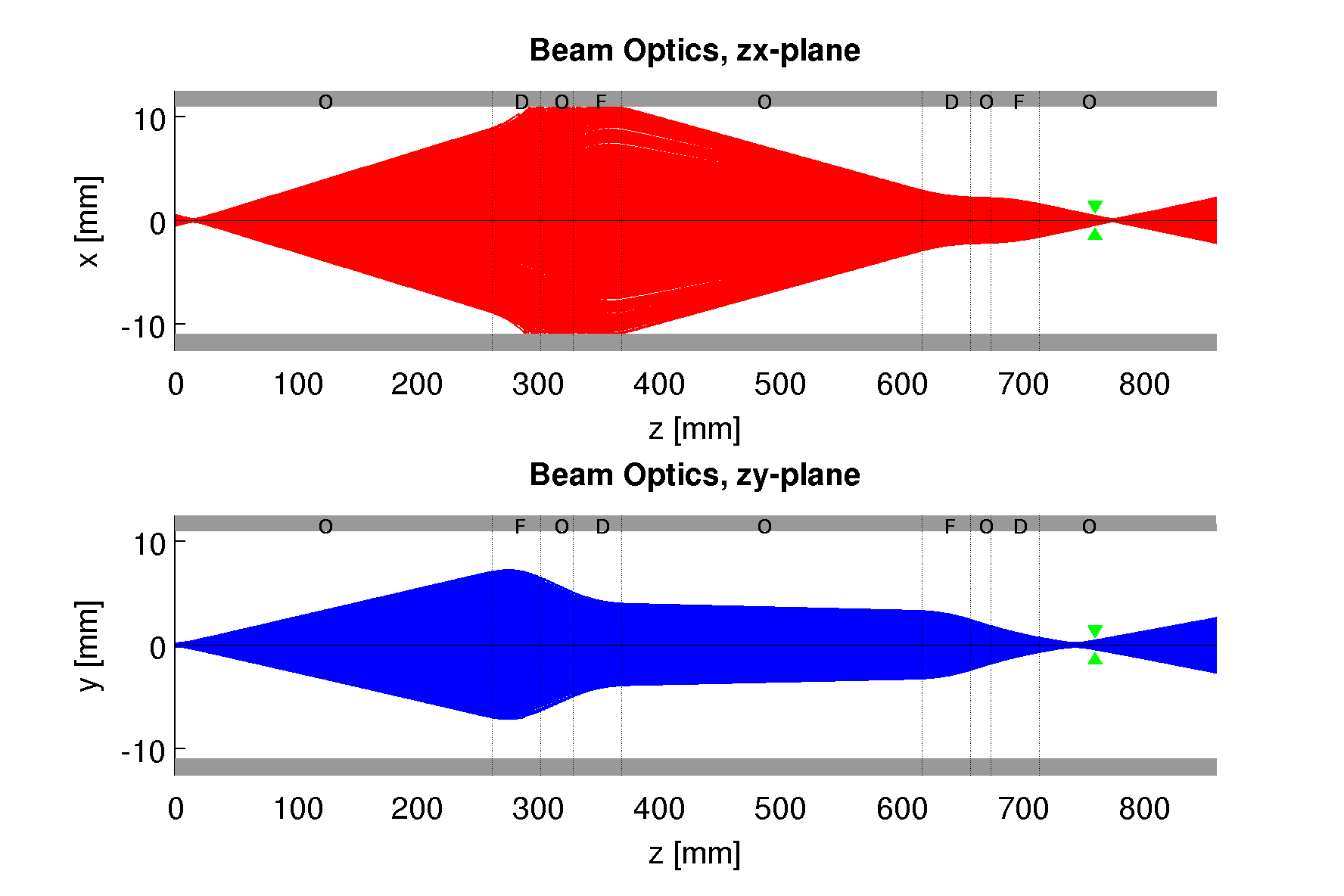}
	\caption[Simulated particle tracks for proposed beam line]{Simulated particle tracks ($10^3$ particles) for the proposed beam line in horizontal (red) and vertical (blue) coordinates. The arrows (green) mark the target position, located near a beam waist in both x- and y-direction. A significant fraction of the beam ($34.7\,\%$) is lost due to the initial defocusing in x-coordinate.}
	\label{fodo_optimum}
\end{figure}

The drift parameter set had to be chosen within certain boundary conditions ($d_1 \ge 200\,\mathrm{mm}$; $d_i \ge 20\,\mathrm{mm},\,i \ne 1$) and optimised. Using an iterative algorithm, a number of local optima were found, one of which was the parameter set $[262\;27\;248\;17\;46]\;$mm, resulting in a spot size of $(0.92 \times 0.88)\;\mathrm{mm^2}$. The simulated particle tracks for this setup are shown in Fig.~\ref{fodo_optimum}. A fraction of the beam is lost to the beam pipe due to the initial defocusing in the x-coordinate. This cannot be completely avoided due to the size of the beam pipe and the minimal distance at which the first PMQ can be positioned in our experimental setup. A possibility for reducing beam losses by a few percent would be the construction of two DOFO doublets, seeing as the beam divergence is slightly larger in the horizontal coordinate (see Fig.~\ref{emittance}). This could easily be achieved without realigning the quadrupoles by rotating the entire beam pipe by $90^{\circ}$. Because the requirements for beam spot size depend highly on the envisioned applications of the ion beam, a minimisation of beam losses is in general not essential.

\subsection{Construction of the PMQs}
Following the results of the simulations, a prototype of the designed PMQ was constructed at the Laboratory for High Energy Physics (LHEP) in Bern. For easier machinability and assembly, the aluminium supporting frame consists of sixteen semicircle plates of 5 mm thickness. These are assembled on a steel test piece of the same dimensions as the beam pipe ({\O} = 25 mm), held in place by four stainless steel aglets. The NdFeB magnets can then be inserted. Finally, the ``lids'' are assembled, again consisting of four individual semicircle pieces. These are fixed with four stainless steel screws, making it impossible for the magnets to move within the frame (Fig.~\ref{pmq_tech}).

\clearpage{}
\section{Measurements and results}

\subsection{Magnetic field in the PMQs}
\begin{figure}[!ht]
	\centering
	\subfloat[ ]{
	\includegraphics[width=0.46\textwidth, trim=0cm 0cm 1cm 0cm, clip=true]{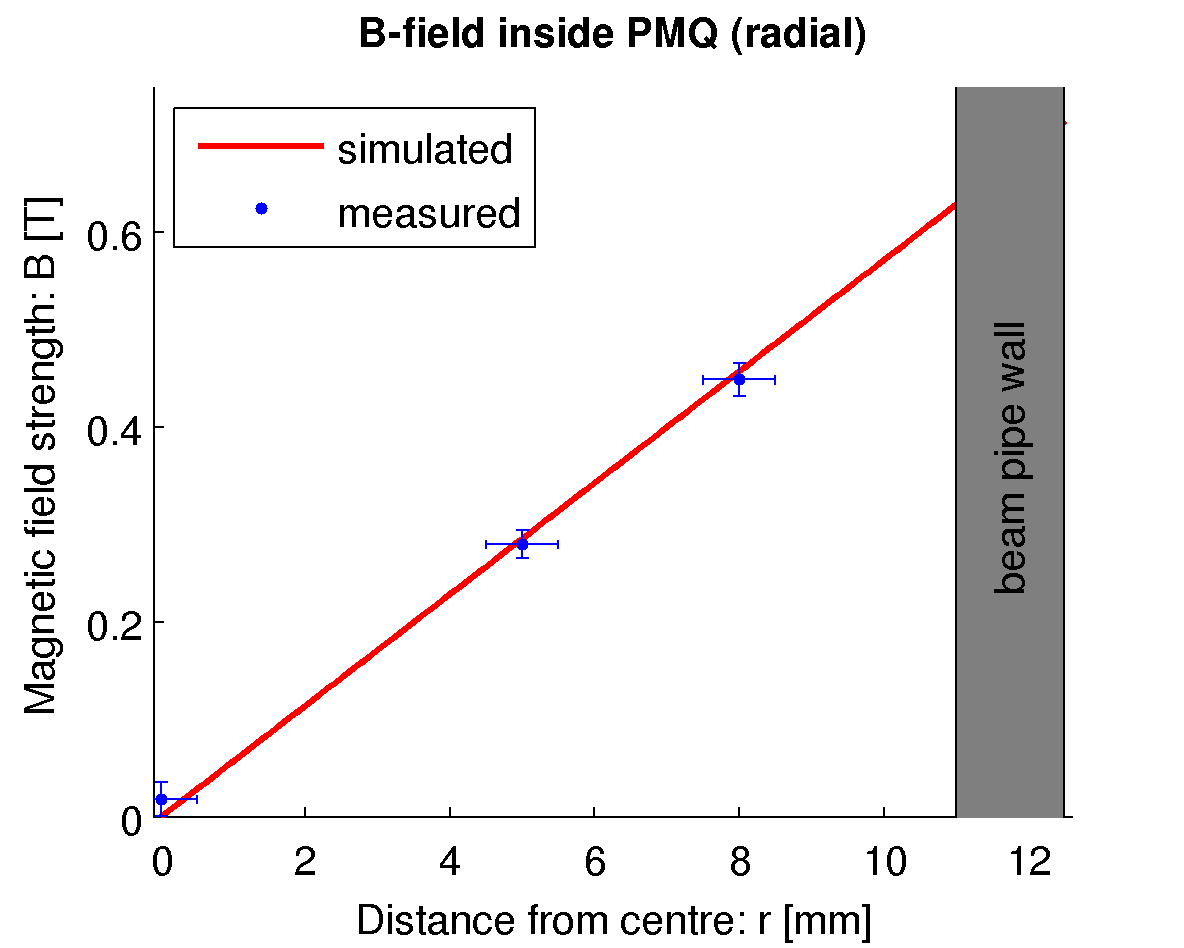}
	\label{B_radial}}
	\qquad
	\subfloat[ ]{
	\includegraphics[width=0.46\textwidth, trim=0cm 0cm 1cm 0cm, clip=true]{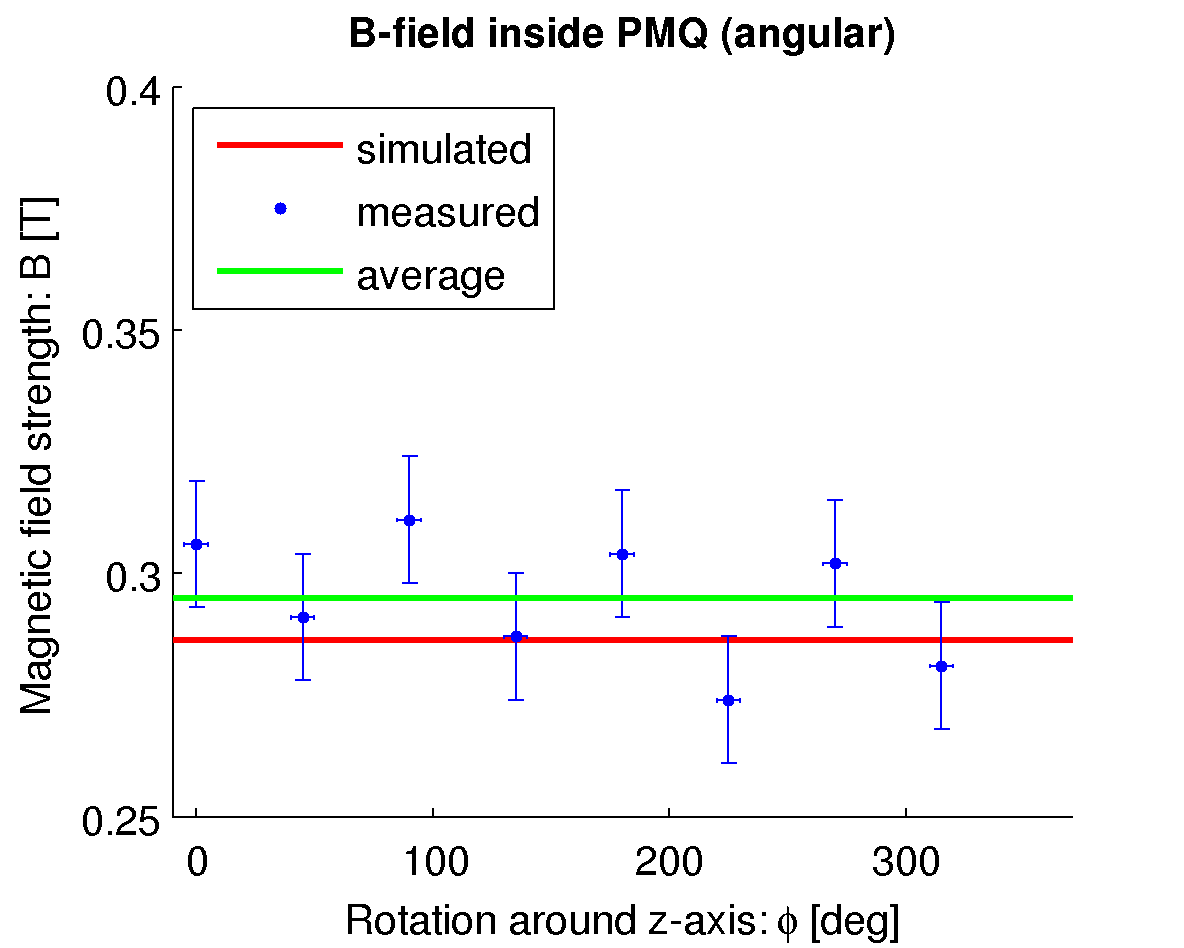}
	\label{B_angular}}
	\caption{Comparison of measured to simulated values for the magnetic field strength within the PMQ. (a) Radial field dependency $B(r)$, averaged measurements at $r=0,\,5,\,8\,\mathrm{mm}$ compared to the simulated field; errors are estimated for $r$ and the standard deviation is used for $B$. (b) Angular field dependency $B(\phi)$, measured values at $r=5\,\mathrm{mm}$ for $\phi=(n-1)\frac{\pi}{4},\;n=1\,...\,8$ compared to the simulated field; errors are estimated for $\phi$ and the standard deviation is used for $B$.}
	\label{B-fields_pmq}
\end{figure}

To confirm the validity of the magnetic field simulations, the PMQ was mounted on a test piece with the same diameter as the beam pipe of the RFQ linac. The magnetic field strength was then measured at various transverse positions at the longitudinal centre of the PMQ using a Hall probe\footnote{LakeShore, S/N H02771, cal. no. 974}. Results are presented in Fig.~\ref{B-fields_pmq}.

For the radial dependency, the measurements were consistent with the simulations with the simulated line lying well within the error bars of the measurement points (Fig.~\ref{B_radial}). For the angular dependency, slightly higher field strengths were found close to those magnets orientated towards the centre of the PMQ (Fig.~\ref{B_angular}). This is to be expected, as a Halbach cylinder does not generate an ideal quadrupole field \cite{halbach0}. This modulation may be due to a slight misalignment of either the PMQ or the Hall probe, or to non-linear field errors. To study these effects, a full 3-D mapping of the magnetic fields within each constructed quadrupole  would be necessary. Within the scope of this work, a simplified model is used: The magnetic field strength is less homogeneous close to the magnets, but becomes more homogeneous near the transversal centre of the PMQ, which is why the magnetic field is later assumed to be angle-independent with a maximal field gradient of $G_{\mathrm{max}}=57$~T/m at the longitudinal centre of the PMQ.

\subsection{Tuning the beam line}
\begin{figure}[!ht]
	\centering
	\includegraphics[width=\textwidth, trim=30cm 35cm 0cm 30cm, clip=true]{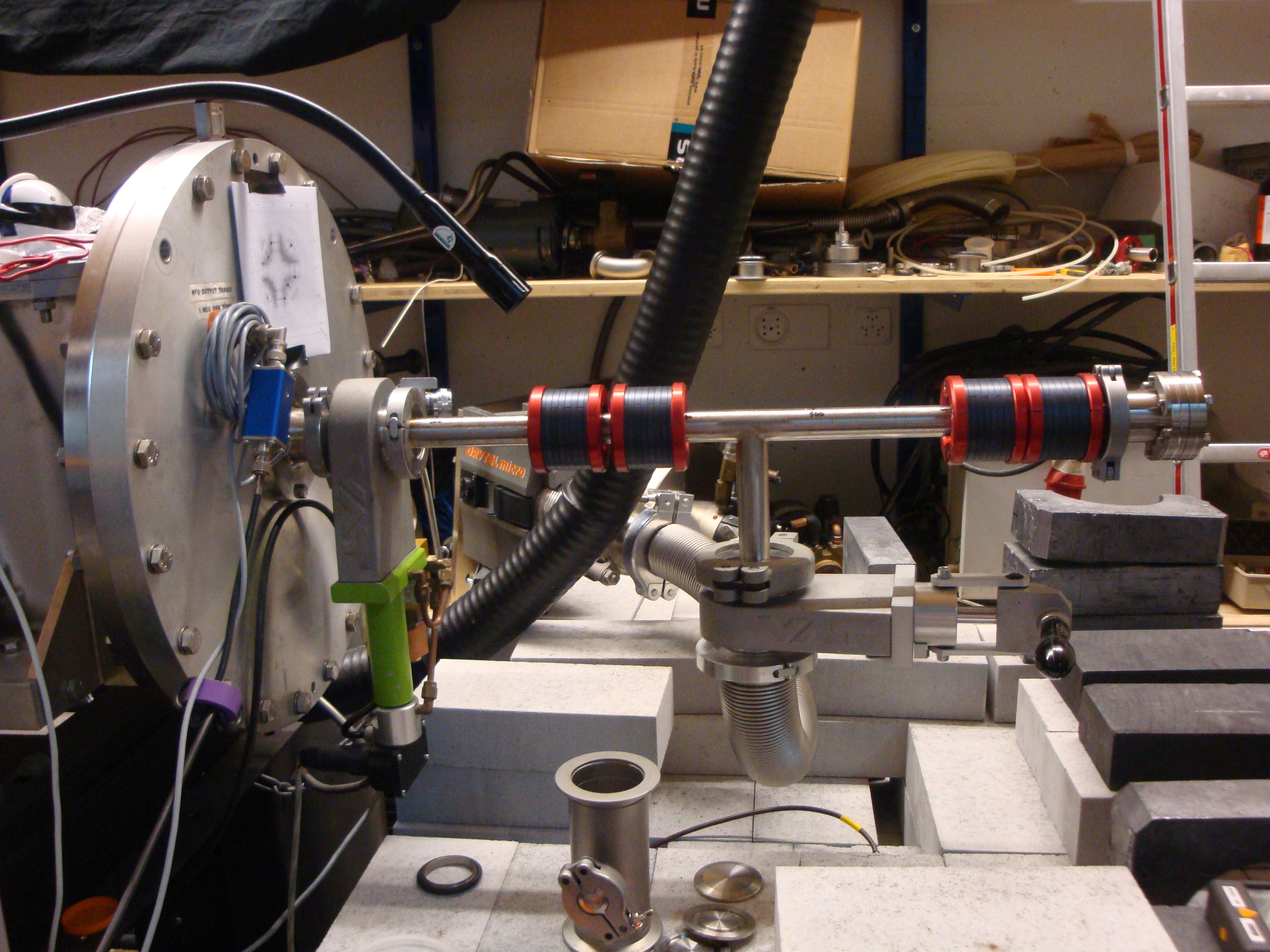}
	\caption{Assembled HEBT beam line of the RFQ Linac with four PMQs.}
	\label{hebt_photo}
\end{figure}

After successfully testing the magnetic field of the PMQ prototype, four additional PMQs were manufactured. The field strength and homogeneity of 50 permanent magnets were measured using the Hall probe. On this basis, a selection of 8 similar magnets was made for each PMQ, in order to minimise deviations from an ideal quadrupole field \cite{halbach0}. The PMQs were arranged into two FODO cells on a stainless steel beam pipe at the positions calculated in the previous sections, as shown in Fig.~\ref{hebt_photo}. The longitudinal position of each PMQ can be varied manually by sliding them along and turning them around the pipe axis. The effect of a certain PMQ's position and orientation on the beam can be observed in real-time on a scintillating glass plate (located inside the beam pipe) through a borosilicate window.

In order to better tune the beam line, a simple device was constructed for moving the glass plate along the beam line without breaking the vacuum. The plate has a thickness of 1 mm and a diameter of 20 mm, and is fixed within a magnetic holder. The holder can be shifted along the beam line by dragging it with a small magnet on the outside of the beam pipe. No effect on the beam could be seen due to this additional magnet.

\subsection{Beam spot size at target position}
\begin{figure}[!ht]
	\centering
	\includegraphics[width=\textwidth, trim=1.5cm 0.5cm 2cm 0.5cm, clip=true]{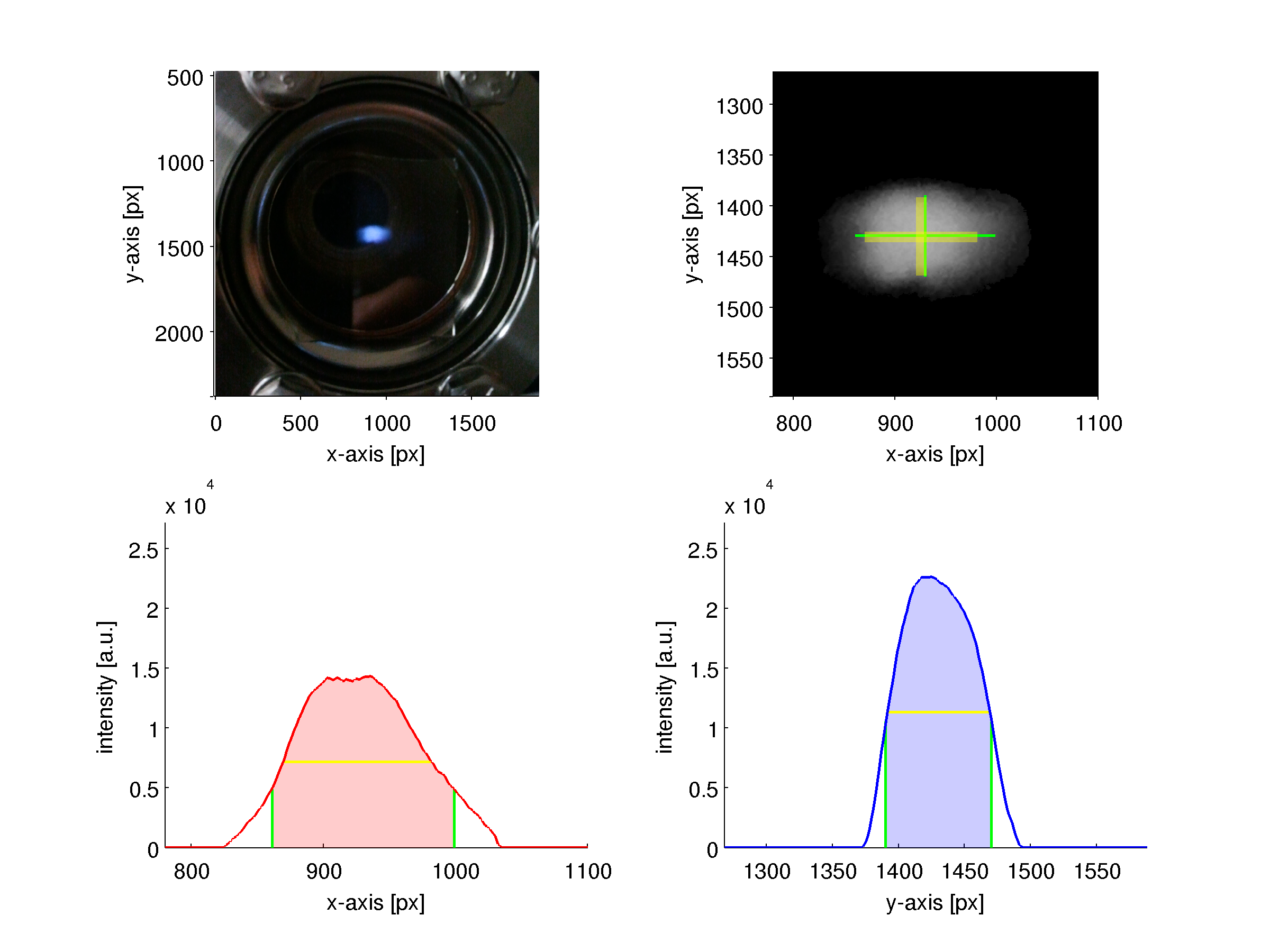}
	\caption{Estimation of the beam spot size. Original image (above left, units are pixels), processed image after background subtraction (above right), and projected intensity profiles onto x- and y-axes (below). The camera resolution is $\sim15$~px/mm. The estimated spot sizes using the FWHM (yellow) and the 90\%INT method (green) are shown.}
	\label{beam_size}
\end{figure}

The size of the beam was estimated using photographs of the glass plate taken with a camera at fixed position. The beam ``width'' and ``height'' were estimated via the projected beam intensity profile onto the x and y axis. This was done using the FWHM of the intensity profile, as well as the width containing 90\% of the integrated intensity profile (indicated as 90\%INT). An example for the measured spot size is given in Fig.~\ref{beam_size}. An absolute uncertainty $\Delta x_i = \Delta y_i = 1$~mm was assumed for the measured beam spot sizes.

\begin{figure}[!ht]
	\centering
	\includegraphics[width=\textwidth, trim=0cm 0cm 0cm 0cm, clip=true]{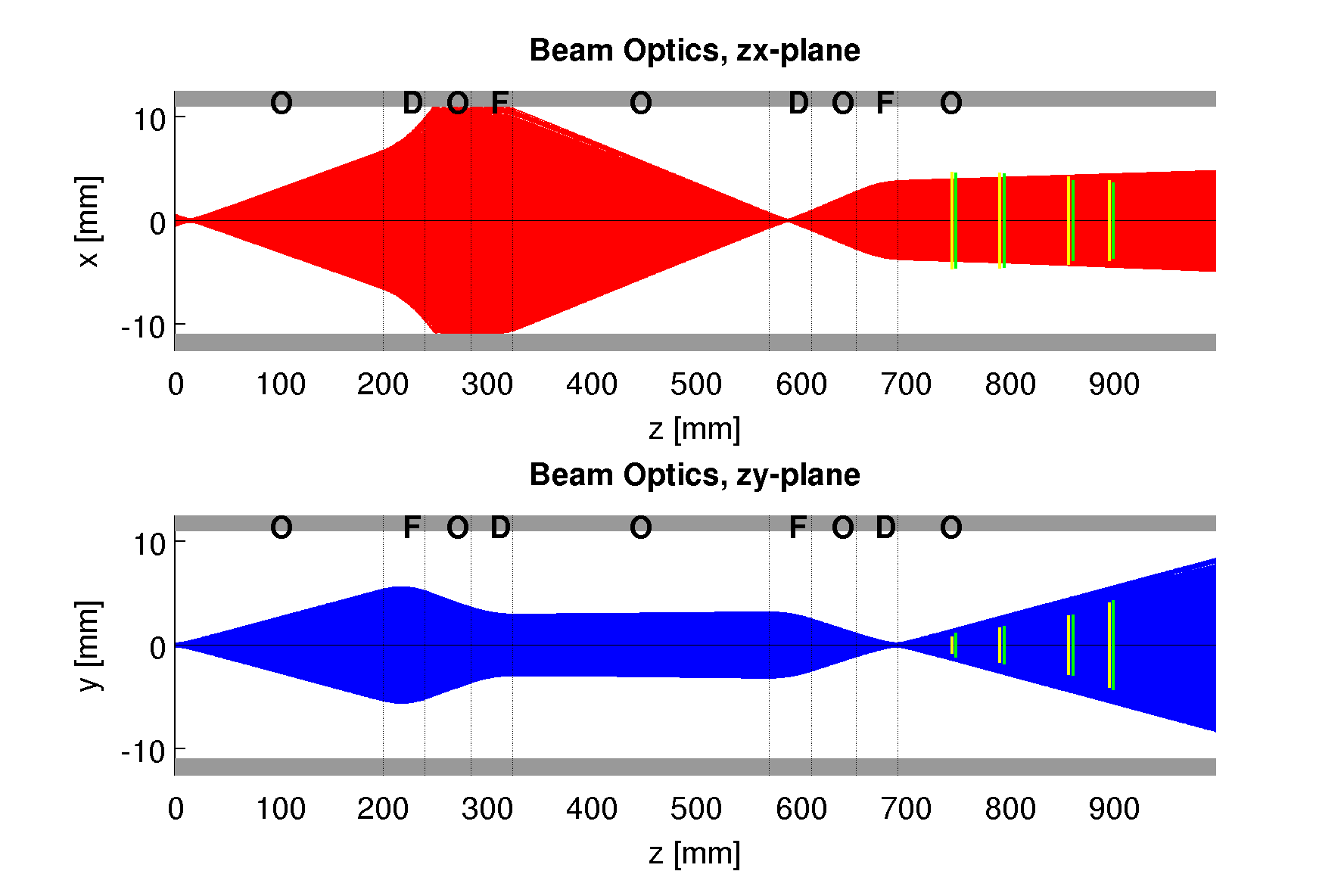}
	\caption{Beam line using the drift configuration $[200\;44\;246\;43\;54/100/166/205]\;$mm. The simulated beam is shown in the horizontal (red) and vertical (blue) plane and compared to the observed spot sizes, where the FWHM (yellow) and 90\%INT (green) estimations are shown. The beam is practically parallel horizontally, but divergent vertically.}
	\label{config_02}
\end{figure}

\begin{figure}[!ht]
	\centering
	\includegraphics[width=\textwidth, trim=0cm 0cm 0cm 0cm, clip=true]{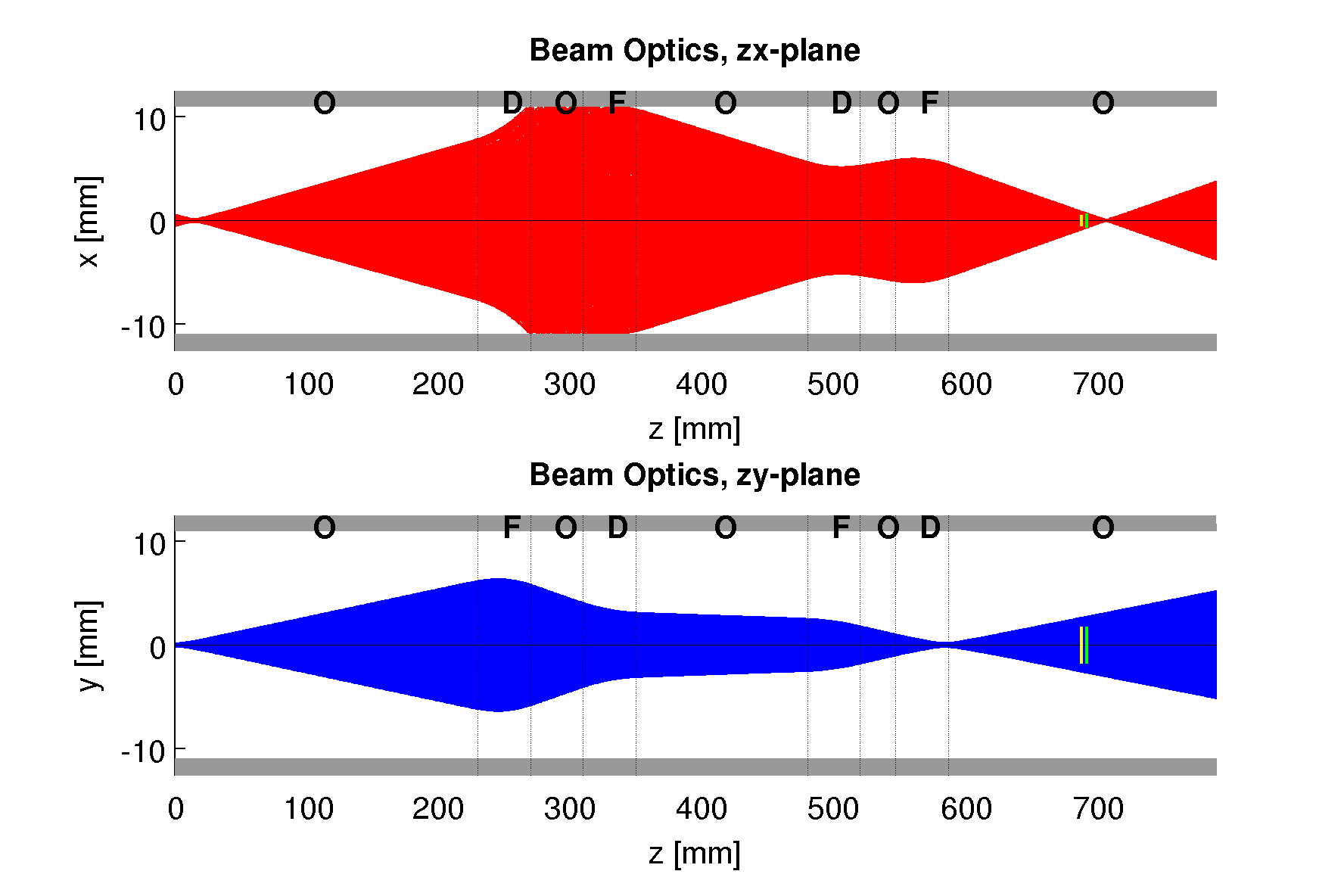}
	\caption{Beam line using the drift configuration $[230\;40\;130\;27\;103]\;$mm. The simulated beam is shown in the horizontal (red) and vertical (blue) plane and compared to the observed spot sizes, where the FWHM (yellow) and 90\%INT (green) estimations are shown. The glass plate is positioned near a beam waist in x-direction; the waist in y-direction can be found inside the last quadrupole.}
	\label{config_12}
\end{figure}

The beam spot sizes were measured for 12 different PMQ configurations at 2-3 target positions on average, yielding a total of $N=30$ measured beam spots. The comparison between simulation $x^{\mathrm{sim}}_i$ and measurements $x^{\mathrm{meas}}_i$ (using the 90\%INT beam size) was done using a quality figure of merit that was defined as follows:
\begin{equation} 
    \chi^2_x = \frac{1}{N} \sum \limits_{i=0}^N \left( \frac{x^{\mathrm{sim}}_i - x^{\mathrm{meas}}_i}{\Delta x_i} \right)^2 .
    \label{chisquare_formula}
\end{equation} 
The average values were found to be $\bar{\chi}^2_x = 3.35$, $\bar{\chi}^2_y = 5.27$. 
The average simulated beam loss in the horizontal plane was 30\%, though the true value is expected to be lower as the beam intensity is not equally distributed within the 90\% emittance. Examples for the agreement between the simulated and the measured beams are given in Fig.~\ref{config_02} and Fig.~\ref{config_12}. A quantitative comparison is shown in Table~\ref{chisquare_table}. For all measurements, we observed that $\chi^2_x < \chi^2_y$; this is most likely caused by the beam collimation in x-coordinate on the beam pipe wall, thereby removing contributions of the beam which lie outside the 90\% emittance.

\begin{figure}[!ht]
	\centering
	\includegraphics[width=0.8\textwidth, trim=0cm 0cm 0cm 0cm, clip=true]{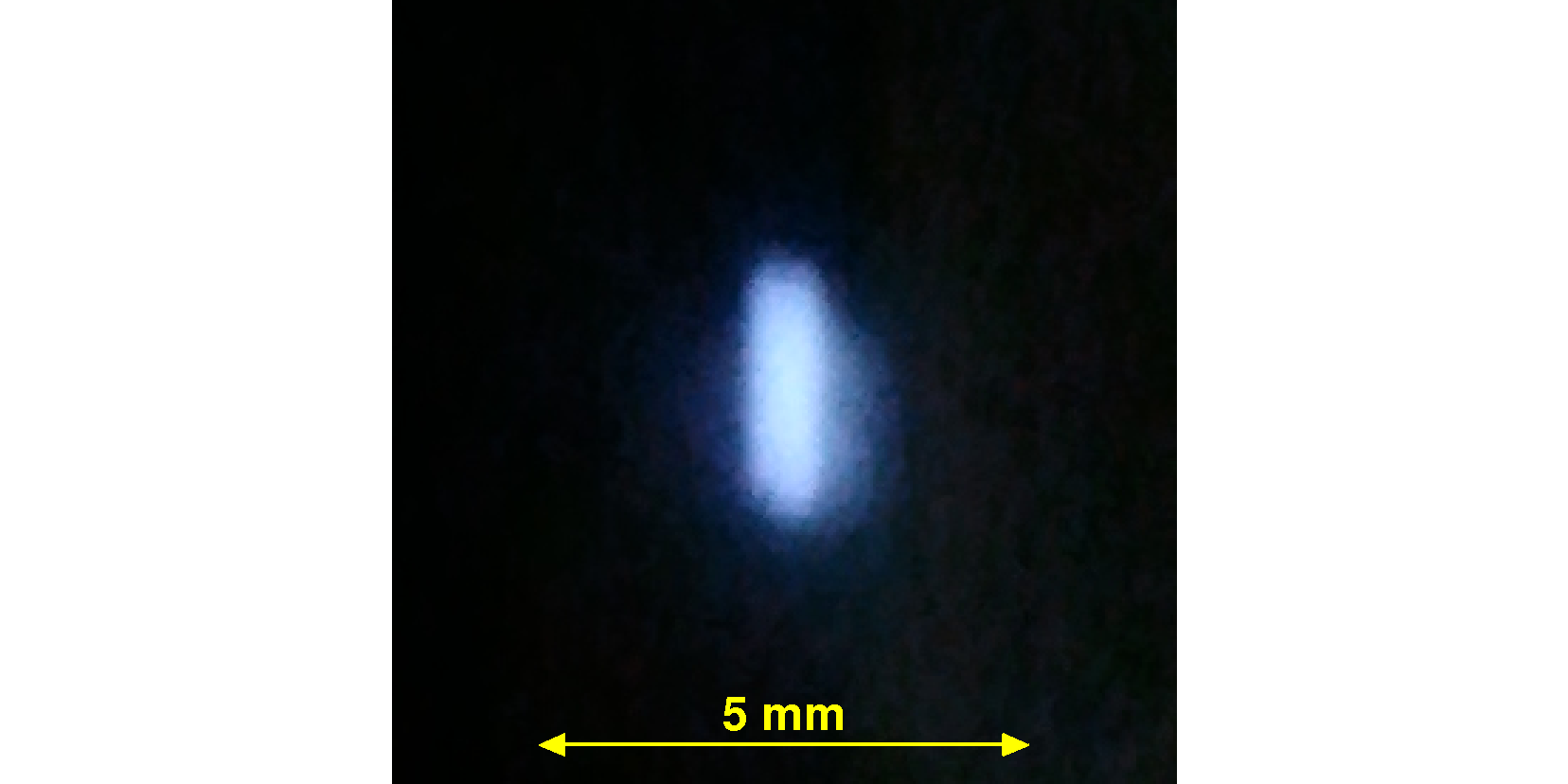}
	\caption{Smallest achieved beam spot size, observed on the scintillating glass screen target.}
	\label{focused_meas}
\end{figure}

Finally, the PMQ positions were adjusted iteratively to tune the beam line, reaching a minimal beam spot for the drift distance parameter set $[229\;21\;159\;20\;203]$~mm (shown in Fig.~\ref{hebt_photo}). The final beam size on target was $(1.2\times2.2)\;\mathrm{mm^2}$, shown in Fig.~\ref{focused_meas}. The particle intensity was estimated using the integrated beam current and the simulated beam losses, resulting in a value of $2.9\cdot10^{14}\;\mathrm{cm^{-2}\,s^{-1}}$. It should be noted that the obtained minimal beam spot size is a factor 3 higher than the simulated optimum, this is likely due to non-linear contributions to the magnetic field which may limit the achievable minimal spot size. Further optimisation of the beam line is possible, should a specific application require smaller spot sizes.
\begin{table}[!ht]
	\begin{center}
		\caption{Comparison of simulation and measurements using the quality figure of merit $\chi^2$.}
		\label{chisquare_table}
		\begin{tabular}{| c | l l l l l | c | c c |}
			\hline
			Fig.~& $d_0$ & $d_1$ & $d_2$ & $d_3$ & $d_4$ [mm] & weight & $\chi^2_x$ & $\chi^2_y$ \\
			\hline
			\ref{config_02} & 200 & 44 & 246 & 43 & 54/100/166/205   &  4 & 1.51 & 5.48 \\
			\ref{config_12} & 230 & 40 & 130 & 27 & 103              &  1 & 0.05 & 3.39 \\
			\hline
			---   & \multicolumn{5}{|l|}{Total for 30 measurements}  & 30 & 3.35 & 5.27 \\
			\hline
		\end{tabular}
	\end{center}
\end{table}

\clearpage{}
\section{Conclusions}
The scope of this work was the realisation of a beam line for a 2 MeV H\textsuperscript{--} ion beam to enable fixed target experiments that require small spot sizes. The high energy beam transport section for an RFQ Linac was realised by means of four permanent magnet quadrupoles. 

Eight cuboid NdFeB permanent magnets with a residual magnetism of $1.3\;\mathrm{T}$ were used to construct Halbach cylinders, arranged in a manner to generate quadrupole fields. The magnetic fields inside the PMQs were simulated using Comsol, the obtained values were confirmed experimentally using a Hall probe. The integrated field of the PMQs was found to be $2.35\;\mathrm{T}$ with a maximal field gradient of $57\;\mathrm{T\,m^{-1}}$.

An emittance measurement provided the initial parameters for the beam optics simulations. A Monte-Carlo particle track simulation code using Matlab and iterative optimisation was employed to find ideal PMQ positions for minimal spot sizes. Starting from these parameters, the PMQs were slid along the beam pipe to tune their positions to an optimum in the experimental setup. The minimal achieved spot size was $(1.2\times2.2)\;\mathrm{mm^2}$, resulting in a particle intensity of $2.9\cdot10^{14}\;\mathrm{cm^{-2}\,s^{-1}}$, beam losses included. The beam was made visible on a moveable scintillating glass plate, observable through a vacuum window. The measured spot sizes agreed with the simulation sufficiently well, though the agreement for the x-coordinate was better than for the y-coordinate. This is most likely due to the horizontal collimation of the beam on the beam pipe wall following the first quadrupole.

Though this work concentrated on realising a small beam spot, the same beam line could be adjusted to obtain a parallel beam (minimal divergence) which would be useful in beam transfer lines. To the authors knowledge, this paper describes the first adjustable PMQ beam line for an external ion beam. The novel compact design based on commercially available magnets allows high flexibility for ion beam applications at low cost.

\acknowledgments
The authors would like to acknowledge the LINAC4 team at CERN for the insight gained with regard to emittance measurements.
We would also like to acknowledge the valuable contributions from the LHEP workshop and technical staff.
We are greatly indebted to the ICSC World Laboratory, owner of the RFQ Linac, and to its president A. Zichichi for giving us the possibility to operate this device at our university.

\clearpage{}
\bibliographystyle{JHEP}
\bibliography{thesis}

\end{document}